# Some Considerations on Neutrinos and on the Measurement of their Velocity

Giorgio Giacomelli
University of Bologna and INFN Section of Bologna
giacomelli@bo.infn.it

**Abstract.** In this report are recalled in a simple form some of the main concepts about neutrinos, starting from their discovery and classifying them in the Standard Model of the Microcosm. Then are presented the main natural sources of neutrinos, emphasizing the enormous number of neutrinos in the Universe. Some information on neutrinos produced by nuclear reactors and by particle accelerators are then considered. In the second part is discussed the neutrino beam sent from CERN to Gran Sasso and the measurement of their velocity performed by the OPERA experiment, underlying some errors made, their corrections, and the final results, which represent precision measurements.

## 1. Introduction

The **neutrinos** are completely outside our every day ordinary experience, and among all the objects present in the Universe are the smallest, the most abundant and the most elusive.

They come from all directions of space, and, every second, thousand billions of them traverse us, but their probability of interacting with ordinary matter is extremely small.

The very many neutrinos coming from our Sun arrive during the day from the top and during the night from the bottom.

The neutrinos are part of the family of the *smallest constituents of matter*, have zero electric charge and a very small mass.

The neutrino was "invented" in 1930 by the german theoretical physicist W. Pauli, in a letter sent to a conference, a short time before going to a "dance party". Pauli invented the neutrino to "save " the conservation laws in a radioactive decay of the type: n$\rightarrow$p+e$^-$ +$\bar{\nu}_e$ .

For almost 30 years the neutrino remained a theoretical particle. Then, with a large detector positioned close to a new nuclear american reactor, in 1953 was detected the $\nu_e$ (with relatively few events) [1]. About 10 years later, at an american accelerator, was detected a second type of neutrino, the $\nu_\mu$, different from the first type [2]. Later at other american accelerators was detected a third type of neutrino, the $\nu_\tau$ [3].

In the *Standard Model of the Microcosm* (*SM*, see Fig. 1) it is assumed that neutrino and antineutrino are *Dirac particle and antiparticle* and thus are different. Now some experiments are trying to verify if they could be equal ( they would be *Majorana neutrinos*).

Are there other types of neutrinos? It would seem no, on the basis of the precision measurements performed at the Large Electron Positron collider (LEP) of CERN (in particular from the precise measurement of the width of the $Z^0$ boson). One could ask if there could be massive neutrinos which would have interaction probabilities even smaller (*sterile neutrinos*); few experiments are being proposed at Fermi National Accelerator Laboratory (*FNAL*) close to Chicago, at the European Center for Particle Physics (*CERN*) in Geneva (CH) and in Japan to verify experimentally this possibility.

Neutrinos are the fundamental particles with the smallest mass (but it is not zero), zero electric charge and interact very rarely with ordinary matter. (See Appendix in pag. 10).

An experimental important fact is that neutrinos of one type spontaneously transform in neutrinos of another type (*Neutrino Oscillations*) [4].

Much remains to be understood about neutrinos, in particular we do not know all their physical properties. Besides we do not know their precise role in the Universe, and if and what they could tell us about the Universe.

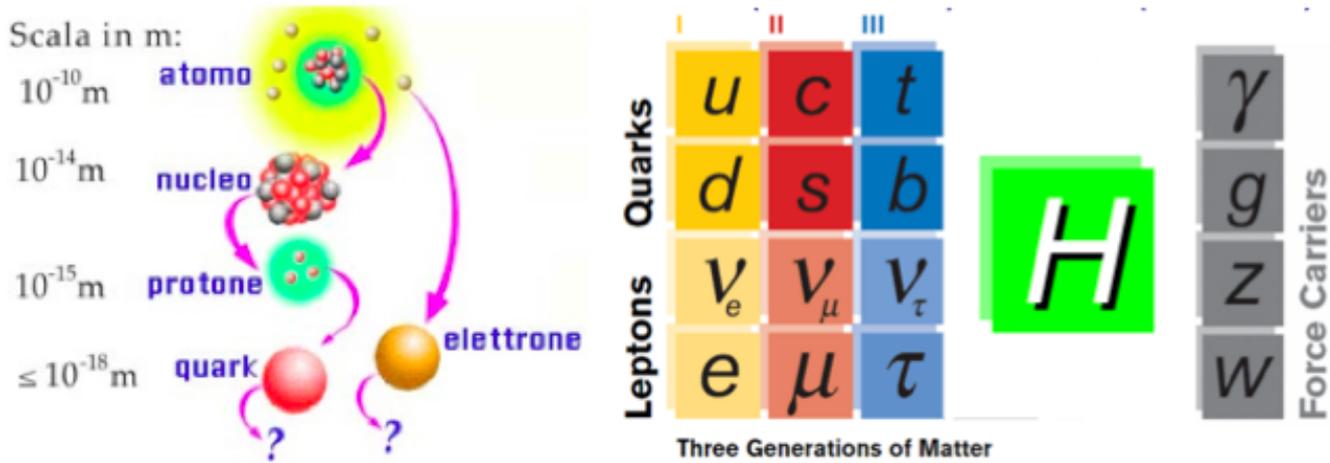

Fig. 1. *The constituents of matter. The fundamental constituents of matter (3 generations), the Higgs boson and the mediators of the fundamental forces (force carriers).*

Fig. 1 left shows the constituents of matter: starting from the molecules and proceeding towards smaller dimensions we find atoms, atomic nuclei with protons and neutrons, the electrons, and the quarks (components of protons and of neutrons). Fig. 1 to the right shows the *fundamental constituents* of matter according to the SM of the Microcosm. They are quarks and leptons, that is *fermions*, that appear in three families: in the first family we find the quark u, d, the electron (e) and the electron neutrino ($\nu_e$). The fundamental constituents of ordinary matter belong to the first family. The fundamental constituents of the second and third family are unstable, seem replicas of the first family and are important only in the context of particle physics. The same figure contains also the carriers of the fundamental forces: of the electromagnetic interaction (the photon $\gamma$ ), of the strong force (the 8 gluons *g*) and of the weak force (the bosons $Z^0$, $W^+$, $W^-$). In the figure is also indicated the Higgs Boson $H^0$, of which has spoken prof. Bellagamba [5].

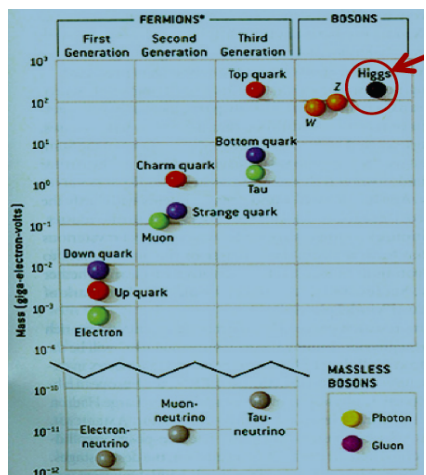

*Fig. 2. Masses of the fundamental constituents, of the force carriers and of the Higgs boson .*

In Fig. 2 are shown all the fundamental constituents, the mediators of the fundamental interactions and of the Higgs boson as a function of their mass. Note the large differences in the mass values of the different particles. The neutrino masses are not well known, but are smaller than 1 eV, while all the other particles have much larger masses. These experimental results find no explanation in the Standard Model of the Microcosm.

We have now to stress that the proton, according to the *static quark model*, is composed by the three almost point like quarks u, u, d, and it seems an almost empty system (Fig. 3a ). But, in the *dynamic quark model* one has to consider completely the strong force among quarks and the proton seems to be full of *virtual particles*, which emerge continuously from the vacuum and remain in the proton for extremely short times, Fig. 3b (thus in pp collisions, one can have for example an interaction of a quark (or a *g*) from the first proton with an anti-quark (or a *g*) of the other proton).

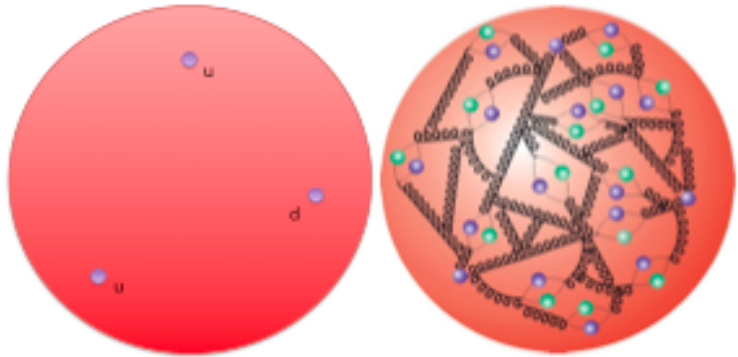

Fig. 3. *(a) The proton according to the static quark model and (b) according to the dynamic quark model.*

## 2. Neutrinos from natural and from artificial sources

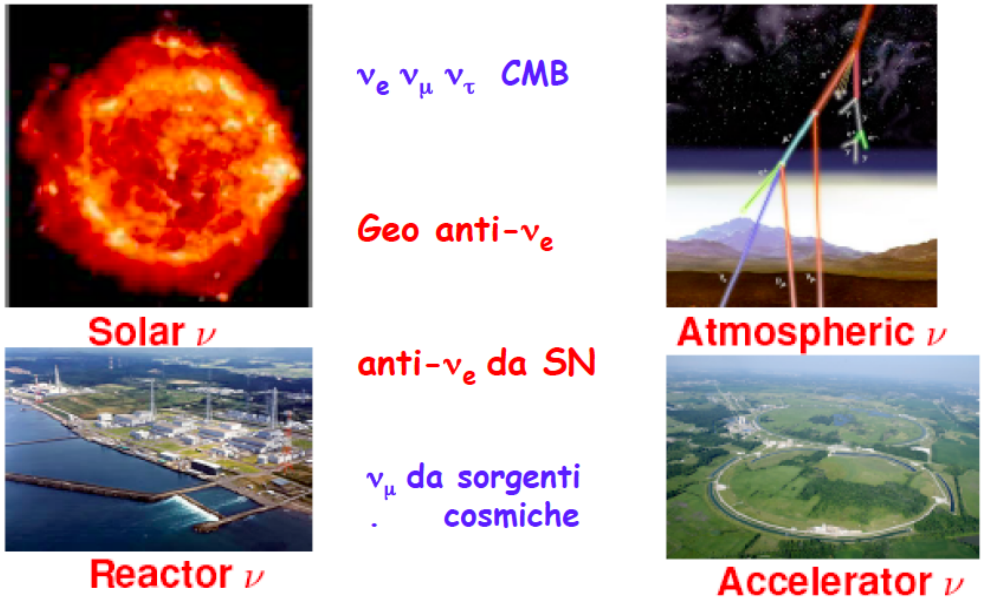

Fig. 4. *General picture of the natural neutrino sources (Sun, Earth, Cosmic Rays, Cosmic Background Radiation, SuperNovae) and from artificial sources (nuclear reactors and particle accelerators ).*

Fig. 4 gives a global picture of the neutrinos produced from natural and artificial sources.

Fig 5 shows the neutrino fluxes which reach the Earth from different natural sources. Note that the main flux comes from the Neutrino Cosmic Background Radiation: we can think that in any point of the Universe, in any instant, one has about 50 ν and 50 anti-ν of each type per cm$^3$, of very low energy: they travel at high speed and practically do not interact at all : it is not possible at this moment to detect them.  A large number of $\nu_e$  are produced by our Sun and they reach the Earth during the day and during the night ($\sim$ 6 10$^{10}$/s cm$^2$); because of neutrino oscillations they arrive on Earth as a mixture of $\nu_e$, $\nu_\mu$, $\nu_\tau$ . The neutrinos coming from the Earth are mainly $\bar{\nu}_e$ from the decays of terrestrial radionuclei with a very long lifetime (from U$^{238}$ and Th$^{232}$). The neutrinos coming from cosmic ray interactions and decays are mainly  $\nu_\mu$ , $\nu_e$ and their anti ν's.

Nuclear reactors produce $\bar{\nu}_e$ (in all directions), while accelerators produce mainly beams of $\nu_\mu$ , very directional. Even if we are in this "sea" of neutrinos, one person in his lifetime, will suffer about 1 interaction of one low energy ( $\sim$1 MeV) neutrino.

Fig. 6 shows the Large Magellano Nebula (a small galaxy at the border of our Via Lattea galaxy). Note the photograph slightly before a star explosion (upper right) and few minutes after the explosion (lower left): the very luminous star at the lower right is the supernova (SN1987A) which emitted a large number of neutrinos: ~20 of them were detected by terrestrial  detectors.

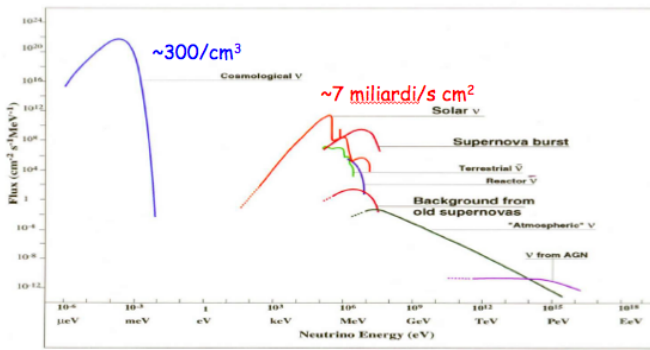 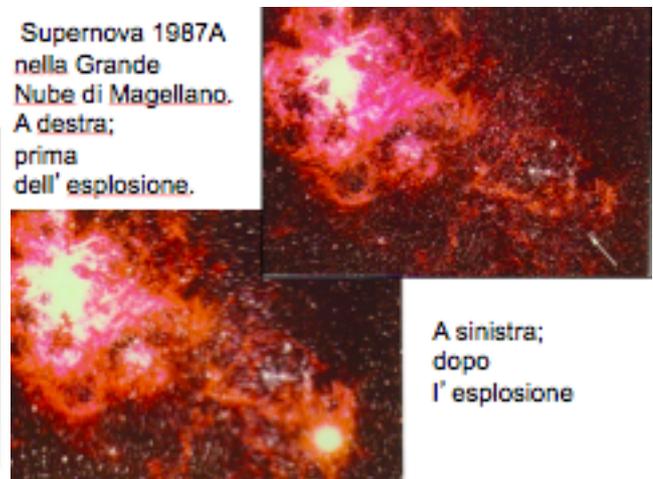

Fig. 5. *Fluxes of  natural ν as function of ν energy.*      Fig. 6. *Supernova 1987A in the Magellano Cloud.*

Recent experiments performed by large international scientific collaborations close to new nuclear reactors (of France-German design) in France (2 reactors), South Korea  (6 reactors) and in China, close to Hong Kong (6 reactors), measured with advanced detectors the last oscillation parameter Θ$_{13}$ for neutrino oscillations, Fig. 7 [6 ] (the Chinese and South Korean centers are those with the largest  world nuclear thermal power).

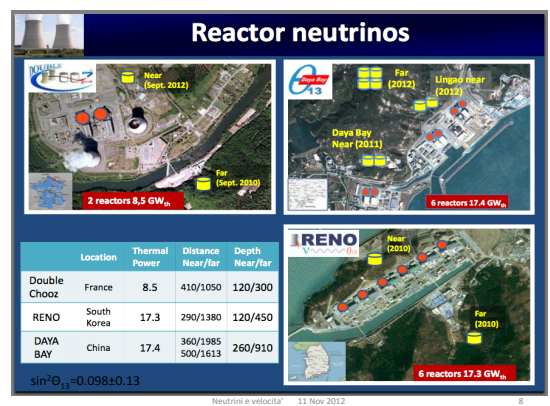

Fig. 7. *Recent experiments with reactor neutrinos performed in France, South Korea and China.*

## 3. Measurement of the muon neutrino velocity in the CNGS beam

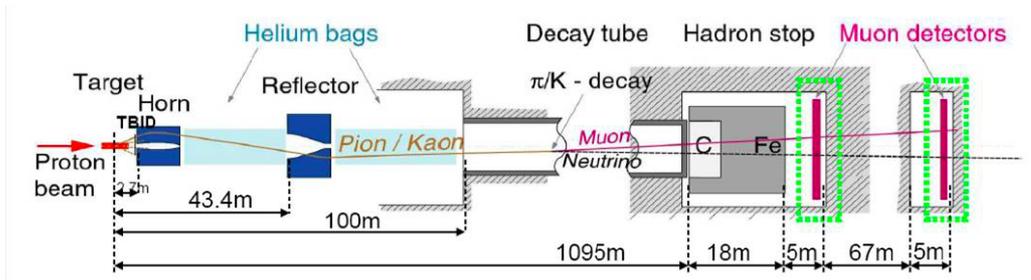

Fig. 8. *CERN : production of the muon neutrino beam from CERN to Gran Sasso (CNGS).*

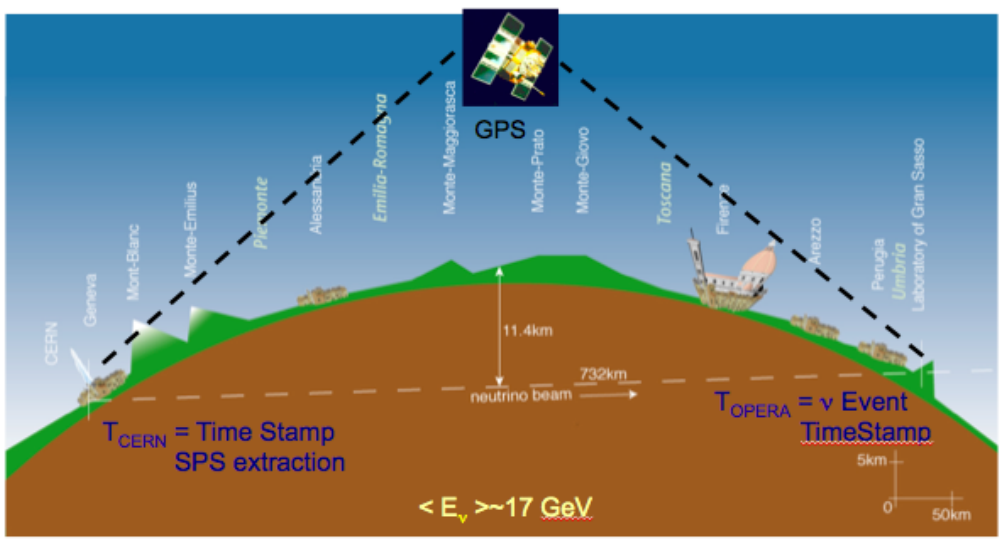

Fig. 9. *The muon neutrino CNGS beam from CERN to Gran Sasso. Note the GPS precision system.*

Fig. 8 shows the structure and complexity of the neutrino beam at CERN using the 450 GeV protons from the SuperSyncrotron (SPS). The protons circulate in the SPS in bunches, which are well controlled, measured in time and intensity before extraction from the SPS and sent in a tunnel where they hit the proton target indicated in Fig. 8, producing charged mesons, which are focused and then decay into muons and muon neutrinos. The muon positions are well measured in the "Muon Detectors". After the measurement of the proton time in the SPS, all other timing fractions of the proton and neutrino trajectories are computed with great precision.

The beam neutrino trajectory from CERN to Gran Sasso is shown in Fig. 9 and in Fig. 10. It is useful to recall that only neutrinos are capable of traversing long distances underground. All other particles stop at the end of the structures shown in Fig. 8. The beam neutrinos have a mean energy of 17 GeV and are $\nu_\mu$ with small contaminations of muon anti-neutrinos and of $\nu_e$.

The measurement of the precise velocity, v, of the $\nu_\mu$ of the CNGS beam became possible after a high precision measurement of the distance CERN-Gran Sasso (with a precision of ±20 cm) and a precision measurement of the time employed to travel that distance (precision ±2 ns), see Fig. 9 and the following formula :

$$v = \frac{\text{CERN} - \text{OPERA path length} = 730 \text{ km} \pm 20 \text{ cm}}{\text{Time Start: protons from SPS} \rightarrow \text{GPS} \rightarrow \text{OPERA Stop} \pm 2 \text{ ns}}$$

The $\nu_\mu$'s were measured in the Gran Sasso underground laboratory by the experiment OPERA [7], composed of two identical modules, each with a target and a magnet, Fig.11 .

The target is made of "bricks" of nuclear emulsions and 1 mm Pb sheets separated by planes of long scintillators horizontal and vertical.  For the measurement of the neutrino velocity are important the scintillators of the Target Trackers (TT): they localize the point where the neutrino interaction occurs and measure the arrival time with a precision of 2 ns. The magnets produce a magnetic field of 1.5 Tesla, thus allowing the measurement, via planes of RPC chambers, of the momentum and of the arrival time of the muon produced in the CC interaction of a $\nu_\mu$ .

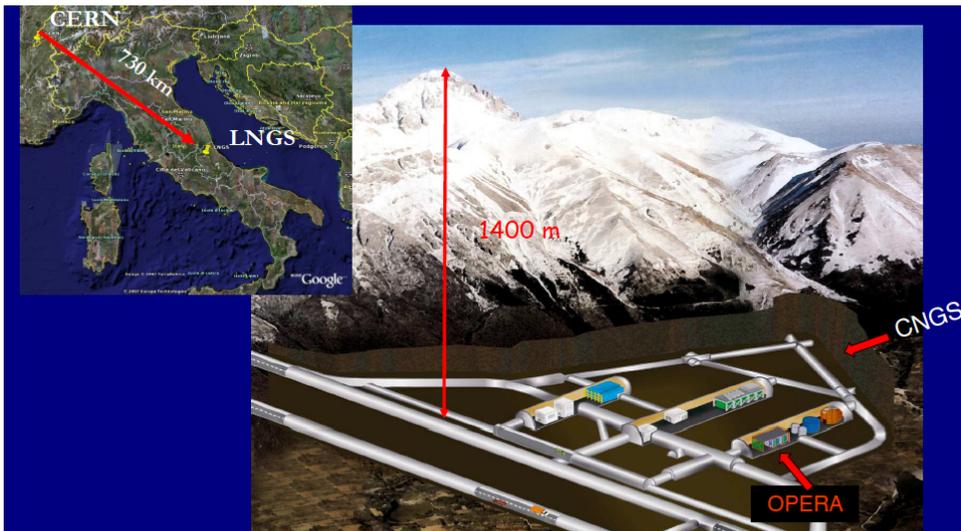

Fig. 10. *Path of the CNGS neutrino beam and scheme of the underground Gran Sasso Lab. (LNGS).*

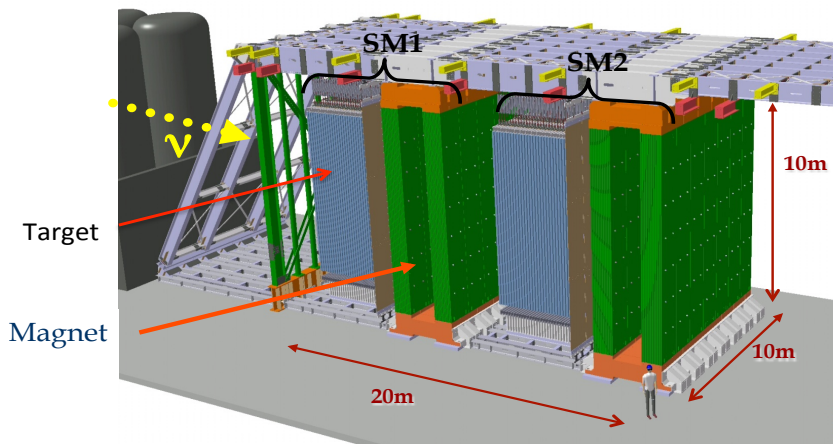

Fig. 11. *The OPERA detector in the underground Gran Sasso lab.*

All the positions at CERN and at Gran Sasso (GS) are measured with great precision using high precision geodetic networks. The information concerning space and time positions are sent from CERN to Gran Sasso via precision GPS, as schematically indicated in Fig. 9. In particular the time of the last passage of the SPS protons is sent to the external Gran Sasso lab. From the external lab this signal is sent to the OPERA experiment via an electrico-optical cable of 8.3 km length; this signal is made in correlation with neutrino candidate interactions in OPERA; in this way one obtains the space and time precisions of 20 cm and of 2 ns.

For the measurements made in 2009-2011 the used neutrino beam was optimized to obtain the highest intensity. In that case the correlation between the SPS proton intensity and the number of neutrino interactions in OPERA could be made only statistically. Later CERN prepared a low intensity neutrino beam, in which the neutrinos were emitted only at specific well defined times as shown in Fig. 14a. In this case it was possible to make a precise measurement between the time of the proton bunch in the SPS and the corresponding arrival time of one neutrino in OPERA.

The measurements made with the first method and with high statistics gave an incredible result: the neutrinos were faster than the speed of light! Naturally this result was very embarassing for all physicists, who immediately suggested a number of checks: remeasure the length of all cables, check the methods of analysis, etc. One of the most important check was to remeasure the delay of the 8.3 km cable which connects the outside Gran Sasso lab to the underground lab. The measurement, that could not be done before the end of the 2011 data taking, was performed on the 6-8 December 2011; it was found that the 8.3 km cable was not properly connected to the OPERA input; this produced a delay of about 73 ns, see Fig. 12. The situation was discussed in a meeting of the authors of the measurement (the Bologna group) one Friday afternoon of December 2011: the calculation of the $\nu_\mu$ velocity based on the measured delay of 73 ns was leading to a $\nu_\mu$ velocity smaller than that of light, as if the neutrinos were heavy particles! This was an incredible un-anticipated result! The calculation was re-checked, obtaining the same result. A strange and chaotic discussion followed, started by a phrase of a young colleague: "let us hope that there is not a second error". In the following discussions the presence of a second, still unknown, error was considered. The situation was summarized saying that a major mistake was found, but the problem was difficult to interpret.

Few days later the Strasbourg group performed a check and found that by sending a signal from a scintillator of the Target Tracker to the data acquisition system it was registered at different times with respect to the used data acquisition cycle. Later they fitted the response and computed an average correction: when this was taken into account it was obtained that the neutrinos were travelling at the speed of light! The second error existed, was localized where we thought it would be, but it was not clear what it was and the effective cause.

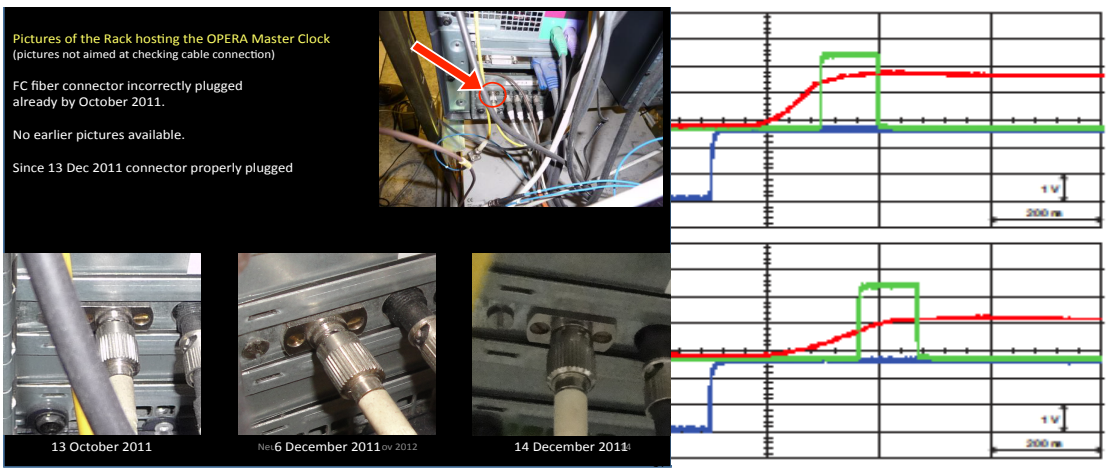

Fig. 12a). *The 8.3 km electro-optical cable arrives to OPERA from the External lab. b) Signals in OPERA: upper graph the cable is properly connected, lower graph the cable is not connected properly: the green signal is delayed by 73 ns.*

At the same time, members of the Bologna group were following the behaviour of few cosmic ray muons which were almost horizontal and passed first in OPERA and then in the experiment LVD (Large Volume Detector) [8], indicated in Fig. 10 by the light blu spot. Because of the low statistics (about 2 muons per week) one had to wait several months: the final situation of the measurements is shown in Fig. 13. The initial correct time situation was that of 2007; then something happened in August 2008, which created an erroneous situation which remained stable until the end of 2011 (it was stable also during the earthquake of 2009). After properly connecting the cable the situation returned equal to the situation of 2007, but all data were obtained in the period 2009-2011 with the incorrect conditions! Also with the first test of the low intensity bunched beam!

Finally with the help and the collaboration of colleagues from Bologna, Strasbourg and Lyon the situation became very clear: the 1st error was due to the capacitance-sensitivity of the connection at the end of the 8.3 km cable; the 2nd error was due to a mis-calibration of the clock in the data acquisition which was recalibrated by colleagues from Bologna and from Lyon using 10 MHz Cs and 5 MHz Rb frequency standards. At this point all errors had been found and were corrected for, see Figs. 12 and 13; immediately followed a re-analysis of the data and finally a correct publication of the results [9].

In May 2012 CERN made a low intensity high quality bunched beam with high time resolution, Fig. 14a. In this case OPERA used the scintillators and the RPC tubes in the magnets, passing through the normal data acquisition system and also using a new system (the Time Board, prepared by the LNF group) which improved the time resolution, see graph in Fig. 14b at the lower right.

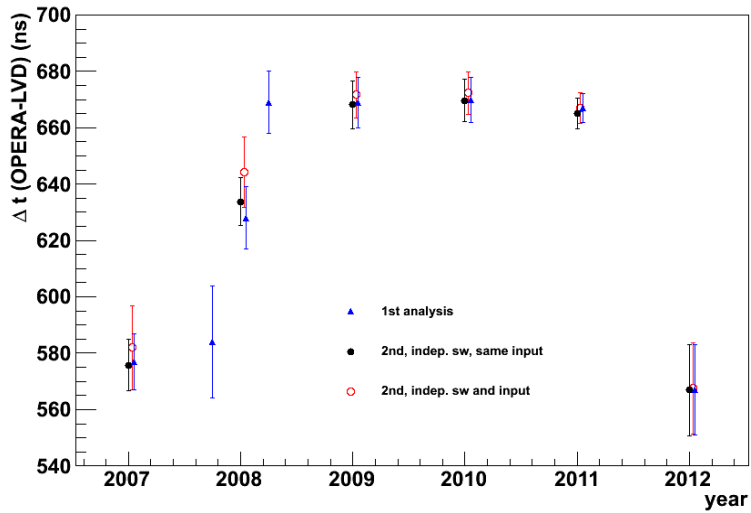

Fig. 13. $\Delta t = t_{LVD} - t_{OPERA}$, as function of the year, for horizontal cosmic ray muons (2nd ind. by Padova and LNF).

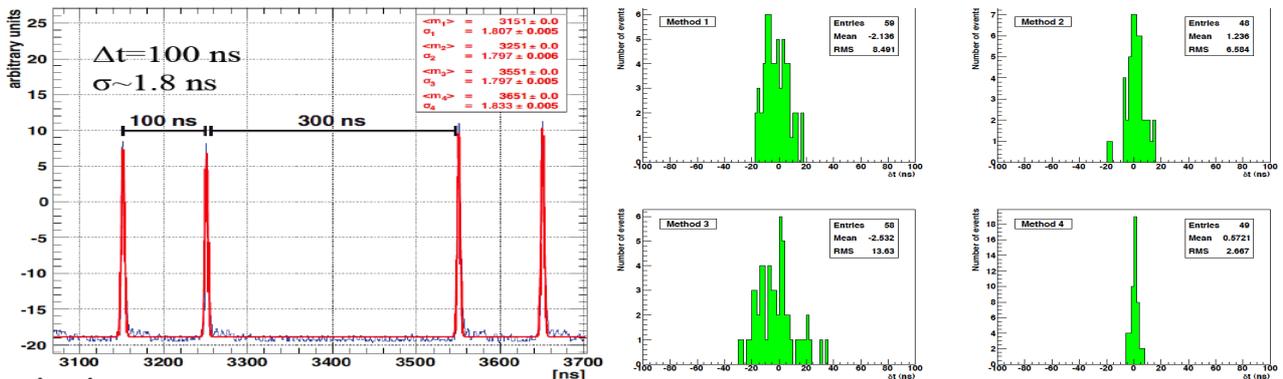

Fig. 14 a) *The CNGS neutrino "bunched beam" in 2012; b) The OPERA results obtained with this beam.*

## 4. Conclusions.

After many difficulties, OPERA with the normal neutrino beam and the re-analisis of 2012, obtained a very good measurement of the 17 GeV muon neutrino velocity v [9]:

$$(v-c)/c = (2.7 \pm 3.1)(stat) \pm 3.4 (sys.) \times 10^{-6}$$

This result is almost one order of magnitude better than the published values of previous measurements.

The measurement made in May 2012 with the special "bunched beam" plus various technical improvements allowed the measurement of the velocity of muon neutrinos [10]

$$-1.8 \times 10^{-6} < (v-c)/c < 3.0 \times 10^{-6} \quad \text{for } \nu_\mu$$

and of the few muon anti-neutrinos present in the bunched beam :

$$-1.6 \times 10^{-6} < (v-c)/c < 3.0 \times 10^{-6} \text{ for anti-}\nu_\mu$$

Now all the OPERA data yield a velocity value for the 17 GeV neutrinos equal to the light velocity! With the new bunched beam other experiments at GS obtained similar results [11].

**Acknowledgements**. I thank the colleagues who organized the popularization of science meeting of Sunday 11 November 2012, in particular profs. G. Dragoni and L. Gregorini. I also thank all the students for the many pertinent questions asked. I acknowledge the colleagues of the OPERA experiment for their collaboration and the solutions of so many problems, obtaining finally a precision result. The acknowledgements are extended to the colleagues of the experiment LVD for the measurement of the delay $t_{LVD}-t_{OPERA}$ and to the CERN colleagues for making optimal CNGS beams.

**Appendix**   (It was asked by the students. Only orders of magnitude values are quoted.)

### -Mass-Energy Units. Orders of magnitude

-1 elettronVolt=1 eV=energy of motion acquired by an electron under a potential difference(ddp)of 1 Volt
-Since motion energy and mass energy are connected by the Einstein relation  **E=mc²**  one can use 1 eV as unit of mass and of energy of motion [ the mass unit may also be written as   $eV/c^2$ ].
  -Some examples:
- $m_\nu \leq 0.1$ eV    ,   or    $m_\nu \leq 0.1$ $eV/c^2$
-1 eV-10 eV  typical energies  of atomic phenomena
-1 keV=$10^3$ eV  maximal energies  of electrons in atoms
- $m_e \sim 0.5$ MeV= 0.5 $10^6$ eV
-1 MeV=$10^6$ eV $\sim$ energy of motion of protons/neutrons in atomic nuclei
-1 MeV – 100 MeV  typical energies of nuclear phenomena studies
-$m_p \sim m_n \sim$ 1 GeV=$10^9$ eV
-$m_{Higgs\ Boson} \sim$ 126 GeV    ;    $m_{quark\ top} \sim$ 180 GeV
-1 TeV=$10^{12}$ eV
-4-7 TeV = energies of motion of protons in LHC

### -Dimensions of the Universe, the Milky Way, the Solar System and of particles in the Microcosm

-Dimension of Universe           $10^{26}$ m
-    "       of Milky Way           $10^{19}$ m
-    "       of Solar System        $10^{12}$ m
-    "         Man                  1 m
-    "         Atoms                $\approx 10^{-10}$ m
-    "         Quarks, Leptons  $\leq 10^{-17}$ m

### -Age, Density, Composition of Universe ; Decoupling of Neutrinos

-Ordinary Matter     4.9 %  from recent Planck measurements
-Dark Matter         26.8 %                "
-Dark Energy         68.3 %                "
-Neutrinos          $\approx$ 0.2 % ?  (Uncertain, they may affect only the very large scale structure in the Universe)
-Age of Universe     13.8 $10^9$ y =13.8 billion years
-Density of Universe $10^{-29}$ g/cm³

-Decoupling of neutrinos in Early Universe    $T_{\nu\ dec} \approx$ 1 MeV

-Mean Present Temperature of Cosmic Background Neutrinos   $T_\nu \approx$ 1.9°K  →  $E_\nu \approx$ 1 eV